\documentclass{webofc}
\usepackage[varg]{txfonts}   % Web of Conferences font
 \usepackage{bm}% bold math

\newcommand{\be}{\begin{equation}}\newcommand{\ee}{\end{equation}}%
\newcommand{\bd}{\begin{displaymath}}\newcommand{\ed}{\end{displaymath}}%%
\newcommand{\bit}{\begin{itemize}}                        %%%%%%%%%
 \newcommand{\eit}{\end{itemize}}                         %%%%%%%%%
\newcommand{\ben}{\begin{enumerate}}                      %%%%%%%%%
 \newcommand{\een}{\end{enumerate}}                       %%%%%%%%%
\newcommand{\baa}{\begin{array}{lll}}                     %%%%%%%%%
 \newcommand{\eaa}{\end{array}}                           %%%%%%%%%
\newcommand{\ba}{\begin{eqnarray}}                        %%%%%%%%%
 \newcommand{\ea}{\end{eqnarray}}                         %%%%%%%%%
\newcommand{\la}{\label}                                  %%%%%%%%%
\newcommand{\A}{{\mathcal{A}}}
\newcommand{\M}{{\mathfrak{A}}}
\newcommand{\I}{{\mathcal{I}}}
\def\1{\hbox{{1}\kern-.25em\hbox{l}}}
\def\Im{\relax{\textbf{Im}{}}}                            %%%%%%%%%
\def\Re{\relax{\textbf{Re}{}}}                            %%%%%%%%%
\newcommand{\Ds}{\displaystyle}                           %%%%%%%%%
\def\as{\relax\ifmmode \alpha_s\else{$ \alpha_s${ }}\fi}  %%%%%%%%%
\def\abar{\relax\ifmmode{\bar{a}}\else{$\bar{a}${ }}\fi}  %%%%%%%%%
\usepackage{color}

   \definecolor{DarkGreen}{rgb}{0.04,0.5,0.1}

\begin{document}

\title{Form factor $\pi^0 \gamma^* \gamma$ in lightcone sum rules
       combined with renormalization-group summation vs experimental data.
       }
%
% subtitle is optional

\author{\firstname{C.} \lastname{Ayala}\inst{1}\fnsep\thanks{\email{cesar.ayala@usm.cl}} \and
        \firstname{S.} \lastname{V. Mikhailov}\inst{2}\fnsep\thanks{\email{mikhs@theor.jinr.ru}} \and
        \firstname{A.} \lastname{V. Pimikov}\inst{3}\fnsep\thanks{\email{pimikov@mail.ru}} \and
        \firstname{N.} \lastname{G. Stefanis}\inst{4}\fnsep\thanks{\email{stefanis@tp2.ruhr-uni-bochum.de}}
        }

\institute{Department of Physics,
           Universidad T\'ecnica Federico Santa Mar\'ia,
           Casilla 110-V, Valpara\'iso, Chile
\and
           Bogoliubov Laboratory of Theoretical Physics, JINR,
           141980 Dubna, Russia
\and
           Institute of Modern Physics, Chinese Academy of Science,
           Lanzhou 730000, China
\and
           Ruhr-Universit\"{a}t Bochum,
           Fakult\"{a}t f\"{u}r Physik and Astronomie,
           Institut f\"{u}r Theoretische Physik II,
           D-44780 Bochum, Germany
          }

\abstract{%
We consider the lightcone sum-rule (LCSR) description of the pion-photon
transition form factor in combination with the renormalization group
of QCD.
The emerging scheme represents a certain version of Fractional
Analytic Perturbation Theory and significantly extends the
applicability domain of perturbation theory towards lower momenta
$Q^2\lesssim 1 $ GeV$^2$.
We show that the predictions calculated herewith agree very well with the
released preliminary data of the BESIII experiment, which have very
small errors just in this region, while the agreement with other data
at higher $Q^2$ is compatible with the LCSR predictions obtained
recently by one of us using fixed-order perturbation theory.
}
\maketitle
\section{Introduction.}
\label{intro}

In this work we consider the calculation of the
$\pi^0\gamma^*\gamma$ transition form factor (TFF)
within the LCSR approach,
see, e.g., \cite{Khodjamirian:1997tk,Mikhailov:2016klg},
going beyond fixed-order perturbation theory (FOPT).
We use instead the approach worked out in \cite{Ayala:2018ifo},
which combines the method of LCSRs, based on dispersion relations,
with the renormalization-group (RG) summation,
expressed in terms of the formal solution of the
Efremov-Radyushkin-Brodsky-Lepage (ERBL)
\cite{Efremov:1978rn,Lepage:1980fj} evolution equation.
We argued \cite{Ayala:2018ifo} that this procedure gives rise to a
particular version of fractional analytic perturbation theory (FAPT)
\cite{Bakulev:2005gw,Bakulev:2006ex}
within QCD---see \cite{Bakulev:2008td,Stefanis:2009kv} for reviews.

From the calculational point of view, this FAPT-related approach
helps avoid the appearance of large radiative corrections to the
pion-photon TFF at low/moderate momenta.
This is because such terms become small by virtue of the FAPT summation
in contrast to the currently known \cite{Mikhailov:2016klg}
FOPT results (up to the order of $O(\alpha_s^2 \beta_0)$).
The emergent FAPT/LCSR approach rearranges completely the perturbative
QCD corrections turning them into a nonpower series of FAPT couplings,
see \cite{Bakulev:2005gw,Bakulev:2006ex,Ayala:2018ifo},
which have no Landau singularities when $Q^2 \simeq \Lambda^2_{\rm QCD}$.
As a result, the domain of applicability of this perturbative expansion
is significantly extended towards lower momentum transfers.
From the phenomenological point of view, the FAPT/LCSR approach extends the validity
range of the TFF predictions below $Q^2\leq 1$~GeV$^2$, where the
preliminary BESIII data on the pion TFF bear very small error bars
\cite{Redmer:2018uew}.
This data regime cannot be reliably assessed using LCSRs within FOPT,
showing a tendency to underestimate them below 1~GeV$^2$
\cite{Stefanis:2019cfn}.

Consider now the pion-photon transition form factor for two highly
virtual photons describing the reaction
$\gamma^*(-Q^2)\gamma^*(-q^2) \to \pi^0$
by assuming that
$Q^2, q^2 \gg m^2_\rho$.
Applying factorization, the pion-photon transition form factor
$F^{\gamma^*\gamma^*\pi^0}$
can be  written in terms of convolutions of perturbatively calculable
hard-scattering parton amplitudes
$T^{(n)}$ of $\gamma^*\gamma^* \to q(G_{\mu\nu})\bar{q}$
and pion distribution amplitudes (DAs) $\varphi_{\pi}^{(n)}$
of nonperturbative nature to get
\begin{eqnarray}
\!\!\!\!\!\!\!\!\!\!\!\!F^{\gamma^*\gamma^*\pi^0}(Q^2,q^2,\mu^2)
\!\!\!&\sim & \!\!\!\!
  T^{(2)}(Q^2,q^2,\mu^2;x)\underset{x}
\otimes
  \varphi_{\pi}^{(2)}(x,\mu^2) +
\label{eq:TFFtw2} \\
&&  \!\!\!\!T^{(4)}(Q^2,q^2,\mu^2;x)\underset{x}
\otimes
  \varphi_{\pi}^{(4)}(x,\mu^2)
  + \text{inverse-power corr. like twist-6}\, ,
\label{eq:TFFtw4-6}
\end{eqnarray}
%Eq (1), (2)
where
$\underset{x}\otimes \equiv \int_0^1 dx$
and the superscript $(n)$ denotes the twist label.
We have adopted the default scale setting by identifying the
factorization (label F) and renormalization (label R) scale
$\mu_\text{F}=\mu_\text{R}=\mu$.
It is possible to sum the infinite series of the logarithmic
corrections associated with the coupling $a_s=\alpha_s(\mu^2)/4\pi$
and the $\varphi_\pi(x;\mu^2)$ renormalization
by absorbing them into the new argument of the running coupling
$\bar{a}_s(q^2\bar{y}+Q^2 y)\equiv \bar{a}_s(y)$
and the ERBL exponent for the DAs, respectively.
For further details we refer to the discussion given in
Sec.\ II of \cite{Ayala:2018ifo} and  references cited therein.
The ERBL exponent accumulates all ERBL evolution kernels $a^{k+1}_s V_k$,
while only the coefficient functions $a^k_s\mathcal{T}^{(k)}$ of the
parton subprocesses remain in the perturbative expansion of
the leading-twist amplitude $T^{(2)}$.

It is useful to consider the twist-two pion DA, as well as the corresponding
contribution to the TFF in (\ref{eq:TFFtw2}), as an expansion in the conformal
basis of the Gegenbauer harmonics
$\{\psi_n(x)=6x\bar{x}C^{3/2}_n(x-\bar{x}) \}$,
\begin{subequations}
\ba
  \varphi_{\pi}^{(2)}(x,\mu^{2})&=& \psi_{0}(x)
  + \sum_{n=2,4, \ldots}^{\infty} a_{n}(\mu^{2}) \psi_{n}(x),
\la{eq:gegen-exp}\\
  F^\text{(tw=2)}(Q^2,q^2)
&=& F_{0}^\text{(tw=2)}(Q^2,q^2)
  + \sum_{n=2,4, \ldots}^{\infty} a_{n}(\mu^{2}) F_{n}^\text{(tw=2)}(Q^2,q^2) \, .\la{eq:gegen-FF}
\ea
\end{subequations}
%Eq (3a), (3b)
At the one-loop level, the next-to-leading-order (NLO) coefficient
function is $\mathcal{T}^{(1)}$ and the RHS of Eq.\ (\ref{eq:TFFtw2})
for the twist-two contribution reduces in the $\{\psi_{n} \}$ basis to
\begin{eqnarray}
\!\!\!\!  F_{n}^\text{(tw=2)}(Q^2,q^2)\stackrel{\text{1-loop}}{\longrightarrow} F_{(1l)n}^\text{(tw=2)}
\!\!\!\!&=&\!\!\!\!\!N_\text{T}T_0(y)
\underset{y}{\otimes}
                           \left\{
                           \left[\1+ \bar{a}_s(y)\mathcal{T}^{(1)}(y,x)\right]
                           \left(\frac{\bar{a}_s(y)}{a_s(\mu^2)} \right)^{\nu_n}
                           \right\}
\underset{x}\otimes
                    \psi_n(x) \, .
\label{eq:T1d}
\end{eqnarray}
%Eq (4)
In the above equation, $T_0(y)$ is the Born term of the perturbative expansion
for $T^{(2)}$, while the other used notations mean
\begin{eqnarray}
&& T_0(y) \equiv T_0(Q^2,q^2;y)
=
  \frac{1}{q^2\bar{y}+Q^2 y}; ~~~ \1
= \delta(x-y); ~~~ N_\text{T}=\sqrt{2}f_\pi/3; \\
&&  V( a_s;y,z) \to a_s V_0(y,z);~~~
  V_0(y,z)\otimes\psi_n(z)=-\frac{1}{2}\gamma_0(n)\psi_n(y);~~~
  \beta(\alpha) \to -a_s^2\beta_0 \, ,
\label{eq:various}
\end{eqnarray}
%Eq (5), (6)
where $a_s\gamma_{0}(n)$ denotes the one-loop anomalous dimension of
the corresponding composite operator of leading twist
with
$\displaystyle\nu_n=\frac{1}{2}\frac{\gamma_0(n)}{\beta_0}$.
It is important to appreciate that Eq.\ (\ref{eq:T1d}) has no sense for
small $q^2$ even if $Q^2$ is large.
Indeed, the scale argument $q^2\bar{y}+Q^2y$ approaches for $y\to 0$
the small $q^2$ regime, so that the perturbative expansion becomes unprotected.
Therefore, Eq.\ (\ref{eq:T1d}) cannot be directly applied to the TFF calculation.
The situation changes drastically when we apply Eq.\ (\ref{eq:T1d}) to a dispersion relation.
\section{Essence of FAPT/LCSRs}
\label{sec-1}
The RG summation of all radiative corrections to the TFF in Eq.\ (\ref{eq:T1d})
generates a new contribution to the imaginary part of the spectral density
(see for details \cite{Ayala:2018ifo})
relative to the standard version of LCSRs
\cite{Khodjamirian:1997tk,Mikhailov:2009kf,Agaev:2010aq,Mikhailov:2016klg}.
Indeed, for the Born contribution the corresponding $\Im$ part is generated by
the singularity of $ T_0(Q^2,-\sigma;y)$ (multiplied by a power of logarithms),
while for the RG summed radiative corrections, one term originates
from the $\Im\left(\bar{a}^\nu_s(-\sigma\bar{y}+Q^2y)/\pi\right)$ contribution.

\subsection{Key element of the radiative corrections}
\label{subsec:key-rad-corr}
The general expression for the key perturbative element follows from the first
term in Eq.\ (\ref{eq:T1d})
\ba
\!\!\!\!\!\!\!\!\!&&T_0(Q^2,q^2;y)\left(\bar{a}^{\nu_n}_s(y)\right) \otimes \psi_n(y)\stackrel{q^2\to -\sigma}{\longrightarrow} \nonumber \\
\!\!\!\!\!\!\!\!\!\!\!\!\!\!\!\!\!\!&& \frac{1}\pi \int_{m^2}^\infty d\sigma
\frac{\Im\big[T_0(Q^2,-\sigma;y)\bar{a}^{\nu_n}_s(-\sigma\bar{y}+Q^2y)\big]}{\sigma+q^2}
\otimes \psi_n(y)= I_n(Q^2,q^2) \label{1} \\
\!\!\!\!\!\!\!\!\!\!\!\!\!\!\!\!\!\!&&=\frac{1}{\pi} \int_{m^2}^\infty \!\frac{d\sigma}{\sigma+q^2}
\left\{\Re[T_0(Q^2,-\sigma;y)]\Im[\bar{a}^{\nu_n}_s(-\sigma\bar{y}+Q^2y)]+  \right. \nonumber \\
&& \left. ~~~~~~~~~~~~~~~~~~~~~~~~~~~~\Im[T_0(Q^2,-\sigma;y)]\Re[\bar{a}^{\nu_n}_s(-\sigma\bar{y}+Q^2y)]
\right\}\otimes \psi_n(y) \nonumber\\
\!\!\!\!\!\!\!\!\!\!\!\!\!\!\!\!\!\!&&=\frac{1}{\pi}\int_{m^2}^\infty \!d\sigma
\frac{\Re[T_0(Q^2,-\sigma;y)]\Im[\bar{a}^{\nu_n}_s(-\sigma\bar{y}+Q^2y)]}{\sigma+q^2}\otimes \psi_n(y)+\bm{0}\otimes \psi_n(y)\, . \!\!\!\!\!\!\!\!\! \label{2}
\ea
%Eq (7), (8)
Now we impose a new condition: We consider the low limit in the dispersion integral on the RHS,
$m^2 \geqslant 0$, to be the threshold of particle production.
This condition affects the outcome of the LCSR even at the level of the Born contribution as we will
discuss shortly and marks a crucial difference from our approach in \cite{Ayala:2018ifo}.
Phenomenologically, $m^2$ can be taken to be $m^2=(2m_\pi)^2 \approx 0.078$ GeV$^2$,
or one can treat it as a fit parameter.
In Eq.\ (\ref{2}) only the first term survives, while the second term vanishes.
After the decomposition of the nominator
$T_0(Q^2,-\sigma;y)\sim 1/(-\sigma \bar{y}+Q^2y)$
and the denominator $\sigma+q^2$ in the integrand and by
replacing there the variables $\sigma \to s= -(- \sigma\bar{y}+Q^2y) \geq 0 $,
one can derive the integral
\begin{subequations}
\label{3}
\ba
\!\!\!\!I_n(Q^2,q^2)&=& -\int_{m(y)}^\infty ds \frac{\rho_{\nu_n}(s)}{s(s+Q(y))} \otimes \psi_n(y) \,,\label{3a}\\
\!\!\!\!\text{where}&&\!\!\!\!\!\!\!\!\!\!\rho_\nu(s)=\frac{1}{\pi}\Im[\bar{a}^\nu_s(-s-i\varepsilon)]; ~Q(y)\equiv q^2\bar{y}+Q^2y; ~m(y)=m^2\bar{y}-Q^2y\, .
\label{3b}
\ea
\end{subequations}
%Eq (9a), Eq (9b)
The value of the low limit $m(y)> 0$ leads to a new constraint for the integration over $s$.
By contrast, taking $m(y)\leqslant 0$, one should start to integrate with $s=0$, where $\rho_{\nu}(s)\neq 0 $.
Hence we have
\ba
\!\!\!\!\!\!\!\!\!I_n(Q^2,q^2)&\!\!\!\!\!=\!\!&\!\!\!\! -\left[\theta(m(y)> 0)\!\int_{m(y)}^\infty ds \frac{\rho_{\nu_n}(s)}{s(s+Q(y))}+
             \theta(m(y)\leqslant 0)\!\int_{0}^\infty\! ds \frac{\rho_{\nu_n}(s)}{s(s+Q(y))}\right]\otimes \psi_n(y)\,, \nonumber\\
\!\!\!\!\! &\!\!\!\!\!=\!\!&\!\!\!\!-\Big[\theta(m(y)> 0) J_{\nu_n}(m(y),Q(y)) +\theta(m(y)\leqslant 0) J_{\nu_n}(0,Q(y))\Big]\otimes \psi_n(y) \,.
           \label{eq:Idecompose}
\ea
%Eq (10)
The new terms $-J_{\nu}$, introduced on the RHS of Eq.\ (\ref{eq:Idecompose}), can be decomposed by means of
the new coupling $\I_{\nu}$ and the previous FAPT couplings $\A_{\nu}, \M_{\nu}$ to obtain
\begin{subequations}
\label{4}
\ba
\!\!\!\!\!\!\!\!\!\!-J_{\nu}(y,x)\!\!\!\!\!\!&=&\!\!\!\!- \int_{y}^\infty ds \frac{\rho_\nu(s)}{s(s+x)}= \frac{1}{x}\left[\I_{\nu}(y,x)- \M_\nu(y)\right] \, ,  \label{4a}\\
\!\!\!\!\!\!\!\!\!\! \I_{\nu}(y,x)\!\!\!\!&\stackrel{\rm def}{=}&\!\!\!\! \int_{y}^\infty \frac{d\sigma}{\sigma+x} \rho_{\nu}^{(l)}(\sigma) \label{4c} \, ,\\
\!\!\!\!\!\!\!\!\!\!\A_{\nu}(x)\!\!\!\!\!&=&\!\!\!\!\!\I_{\nu}(y \to 0,x), \M_{\nu}(y)=\I_{\nu}(y,x \to 0), \I_{1}(y \to 0,x \to 0)=\A_{1}(0)=\M_{1}(0)\, . \label{4d}
\ea
\end{subequations}
%Eq (11a), (11b), (11c)
Substituting Eq.\ (\ref{4a}) into Eq.\ (\ref{eq:Idecompose}),
one arrives at the final expression for $I_n$, notably,
\ba
  I_n(Q^2,q^2)
&=&
  T_0(Q^2,q^2;y)\Big\{\left[\I_{\nu}(m(y),Q(y))-\M_\nu(m(y))\right]
  \theta\left(y <\alpha/(1+\alpha)\right) \nonumber \\
&&\phantom{\!\!\!T_0(Q^2,q^2;y)\Big\{ } \!\!\! + \left[\A_{\nu}(Q(y))-\M_\nu( 0)\right]\theta\left(y \geqslant \alpha/(1+\alpha)\right) \Big\}\otimes \psi_n(y)\, , \label{eq:5}
\ea
%Eq (12)
where $\alpha=m^2/Q^2$
and the former couplings appear as the limit of the expressions
$\I_{\nu}$ in their arguments, cf.\ (\ref{4d}).
Note that the appearance of coupling differences in the square brackets
in Eq.\ (\ref{eq:5}) follows from the decomposition in the integrand on
the RHS of Eq.\ (\ref{4a}).

Turn now to the spectral density.
For this we use the standard FAPT expression for the spectral density $\rho_{\nu}$, i.e.,
 \begin{eqnarray}
 \label{eq:rho-fapt}
\!\!\!\!\!\!  \rho_{\nu}^{(l)}(\sigma)
=
  \frac{1}{\pi}\,
  \textbf{Im}\,\big[a^{\nu}_{(l)}(-\sigma)\big]
= \frac{1}{\pi}\,\frac{\sin[\nu~
  \varphi_{(l)}(\sigma)]}{\left(R_{(l)}(\sigma)\right)^{\nu}} \stackrel{\text{1-loop}}{\longrightarrow} \frac{1}{\pi}\,
   \frac{\sin\left[\nu~\arccos\left(L_{\sigma}/\sqrt{L^2_\sigma+\pi^2}\right)\right] }{\beta_0^\nu~\left[\pi^2+L^2_\sigma\right]^{\nu/2}}\, ,
\label{eq:spec-dens-nu-C2} \nonumber
\end{eqnarray}
where the phase $\varphi_{(l)}$ and the radial part $R_{(l)}$ have
a $l$-loop content, see \cite{Bakulev:2006ex} for details, and
$L_\sigma=\ln(\sigma/\Lambda^2_{\rm QCD})$.
In the equations above, $\M_\nu$ and $\A_\nu$ are the standard FAPT couplings
for the timelike \cite{Bakulev:2006ex} and spacelike \cite{Bakulev:2005gw}
regions, respectively, while the integral $\I_{\nu}(y,x)$ is the new two-parameter
coupling in FAPT, introduced in \cite{Ayala:2018ifo}, and represents a generalization
of the previous FAPT couplings,
 \ba
\!\!\!\!\!\!\!\I_{\nu}(y,x)= \int_{y}^\infty \frac{ds}{s+x} \rho_{\nu}(s)=\A_{\nu}(x) -\!\!  \int_{0}^y \frac{ds}{s+x} \rho_{\nu}(s)=
\M_{\nu}(y) - x\int_{y}^\infty \frac{ds}{s(s+x)} \rho_{\nu}(s)\, . \label{eq:A-def}
\ea
%Eq (13)
For our further considerations it is instructive to define an effective coupling
$\bm{\mathbb{A}_\nu}$ in terms of the parameter
$y_0=m^2/(m^2 +Q^2)$ as follows
\begin{eqnarray}
\label{eq:eff-coupl}
  \!\!\!\!\!\!\!\!\!\!\mathbb{A}_\nu(m^2,y)
 =
  \left[\A_{\nu}(Q(y))-\M_\nu( 0)\right]\theta\left(y\geqslant y_0\right)+ \left[\I_{\nu}(m(y),Q(y))-\M_\nu(m(y))\right]
  \theta\left(y < y_0\right)\,.
\end{eqnarray}
%Eq (14)
Tuning $\alpha$ to larger values, the second term in (\ref{eq:eff-coupl}) becomes more
dominant.
On the other hand, in the vicinity of $y_0$ for $m(y_0)=0$,
$\mathbb{A}_\nu(m^2,y)$ is a continuous function by virtue of the properties (\ref{4d}).
To derive practical results, we use the non-threshold approximation
$\mathbb{A}_\nu(0,y) \to \left[\A_{\nu}(Q(y))-\M_\nu( 0)\right]$
obtained for $m^2 \to 0$.

\subsection{Pion-photon TFF in FAPT}
\label{subsec:TFF-FAPT}
We show the results for the TFF
$F^\text{(tw=2)}_\text{FAPT}(Q^2;m^2)$, obtained from Eq.\ (\ref{eq:T1d}),
by taking into account definition (\ref{eq:eff-coupl}) of the
effective coupling $\bm{\mathbb{A}_\nu}$
in the limits $q^2 \to 0$, $Q(y) \to yQ^2$, and $m^2 \geqslant 0$ in the
following explicit form
\begin{subequations}
\label{eq:TFFq0}
\begin{eqnarray}
\label{n0pTFFq0}
\!\!\!\!\!\!  \nu(n=0)=0;\!\!&\!\!&\!\!\!\!\!\! Q^2 F^\text{(tw=2)}_{\text{FAPT},0}
\equiv
  F_0(Q^2;m^2)
=
  N_\text{T}\left\{\int^1_{0}  \frac{\psi_0(x)}{x}~dx  \right.\nonumber \\
 &&\left. + \left(\frac{\mathbb{A}_{1}(m^2,y)}{y}\right)
\underset{y}{\otimes}
  \mathcal{T}^{(1)}(y,x)
\underset{x}{\otimes}
  \psi_0(x)\right\} \, ,\\
\!\!\!\!\!\!\nu(n\neq0)\neq0;\!\! &\!\!&\!\!\!\!\!\!  Q^2 F^\text{(tw=2)}_{\text{FAPT},n}
\equiv
  F_n(Q^2;m^2)
=
\nonumber \\
\!\!\!\!\!\!\!\!\!\!\!\!\!\!&&\!\!\!\!\!\!\!\!\!\!\!\!\!\!  \frac{N_\text{T}}{a_s^{\nu_n}(\mu^2)}
  \left\{\left(\frac{\mathbb{A}_{\nu_n}(m^2,y)}{y}\right)
\underset{y}{\otimes}
  \psi_n(y)+\left(\frac{\mathbb{A}_{1+{\nu_n}}(m^2,y)}{y}\right)
\underset{y}{\otimes}
  \mathcal{T}^{(1)}(y,x)
\underset{x}{\otimes}
  \psi_n(x)\right\} \, .
\label{npTFFq0}
\end{eqnarray}
\end{subequations}
%Eq (15a), (15b)
These equations can  be related to the initial expressions
given by Eqs.\ (\ref{eq:T1d}) by means of the replacement
$
 \mathbb{A}_{\nu}(m^2,y)
 \rightarrow \bar{a}_s^\nu(y)
$.
The advantage of Eqs.\ (\ref{eq:TFFq0}) is that it does not contain
Landau singularities in $\mathbb{A}_{\nu_n}(0,y)$,
in contrast to Eq.\ (\ref{eq:T1d}), making it possible to integrate over $y$.
As it was discussed in detail in \cite{Ayala:2018ifo}, the singularities of the
FAPT couplings do not disappear completely, but reveal themselves at the end point
$Q^2=0$ for specific values of the index $0 < \nu < 1$.
On the other hand, the magnitude
${\cal A }^{(1)}_{1}(0)={\mathfrak A}^{}_{1}(0)=1/\beta_0$
disturbs the asymptotic value $\sqrt{2}f_\pi$ of the TFF in (\ref{n0pTFFq0}).
Therefore, to save the meaning of the effective coupling
$\bm{\mathbb{A}_\nu}$ in (\ref{eq:eff-coupl}),
we have proposed in \cite{Ayala:2018ifo} ``calibration conditions'' for
${\cal A }^{(1)}_{\nu}(Q^2), {\mathfrak A}^{(1)}_{\nu}(Q^2)$ at the origin
\be
{\cal A }^{}_{\nu}(0)
={\mathfrak A}^{}_{\nu}(0)=0 ~~\text{for} ~~~0< \nu \leqslant 1.
\ee
%Eq (16)
\section{Pion TFF in the FAPT/LCSR approach and comparison with experiment}

\subsection{FAPT/LCSRs for the pion-photon TFF at work}
\label{subsec:FAPT-LCSR-work}
The rearranged perturbative series expansion for the LCSRs via FAPT
leads to the new effective couplings
$\mathbb{A}_\nu(m^2=0,s_0;x)$ (``hard part'') and
$\Delta_{\nu}(m^2=0,x)$ (``resonance part'')
of the LCSRs, where we have taken the limit $m^2=0$.
These effective couplings consist of the same initial FAPT couplings, like in definition
(\ref{eq:eff-coupl}), and possess the same structure, despite the low threshold $m^2=0$.
This should not surprise us, given that the LCSR employs a photon-meson model
that involves only a single parameter, namely, the threshold $s_0$,
i.e., the duality interval for the vector channel,
\begin{eqnarray}
\!\!\!\!\!\!\! \mathbb{A}_\nu(0,s_0;x)
\!\!&=&\!\!
  \theta \left(x\geqslant x_0 \right)
  \left[\A_{\nu}(Q(x))-\M_\nu(0)\right] \nonumber \\
\!\!\!\!\!&&\!\!\!\!\!\!  + \theta\left(x < x_0\right)\left[\I_{\nu}(s_0(x),Q(x))-\M_\nu(s_0(x))\right]\,,
\label{eq:eff-coupl3} \\
\!\!\!\!\!\!\!\!\!\!\!\!
 \mathbb{A}_{\nu}(0;x)-\mathbb{A}_{\nu}(0,s_0;x)
\!\!&=&\!\!\!\! \theta(x < x_0)~\Delta_{\nu}(0,x) \,,\nonumber \\
\!\!\!\!\!\!\!\!\Delta_{\nu}(0,x)&\!\!\!=\!\!\!&\!\!\! \left[\A_{\nu}(Q(x))-\I_{\nu}(s_0(x),Q(x))+\M_\nu(s_0(x))-\M_\nu(0)\right]\,
\label{eq:Aeffect-m=0}\, ,
\end{eqnarray}
%Eq (17), (18)
where  $s_0(x)=s_0\bar{x}-Q^2x$
(in close analogy to Eq.\ (\ref{3b}) for $m(y)$), $x_0=s_0/(s_0+Q^2)$.
We will not derive here the LCSR for the TFF, recommending for further reading
Sec.\ IV in \cite{Ayala:2018ifo}.
We present instead the final results for the partial expressions, see Eq.(\ref{eq:gegen-FF}), pertaining to $F^{\gamma\pi}_{\text{LCSR};n}$
\begin{eqnarray}
\label{eq:37}
 \!\! \!\! F^{\gamma\pi}_\text{LCSR}\left(Q^2\right)
=
 F^{\gamma\pi}_{\text{LCSR};0}\left(Q^2\right)
 + \sum_{n=2,4,\ldots} a_n(\mu^2)~F^{\gamma\pi}_{\text{LCSR};n}\left(Q^2\right),
\end{eqnarray}
%Eq (19)
\begin{subequations}
\label{eq:finalFAPTLCSR}
\begin{eqnarray}
\!\!\!\!\!\!&&\!\!\!\!\!\!Q^2 F^{\gamma\pi}_{\text{LCSR};0}\left(Q^2\right)
\!\!=\!\!
N_{\text{T}}\Bigg\{  \int^{\bar{x}_0}_{0} \!\! \bar{\rho}_0(Q^2,x)\frac{dx}{\bar{x}}+
\!\frac{Q^2}{m_{\rho}^2} \int^{1}_{\bar{x}_0}
  \exp\left(
            \Ds \frac{m_{\rho}^2}{M^2}- \frac{Q^2}{M^2}\frac{\bar{x}}{x}
      \right)\bar{\rho}_0(Q^2,x)\frac{dx}{x} \label{eq:38a} \\
\!\!\!\!\!\!\!\!\!\!&&\!\! +   \left(\frac{\mathbb{A}_{1}(0,s_0;x)}{x}\right)
\underset{x}{\otimes}
  \mathcal{T}^{(1)}(x,y)
\underset{y}{\otimes}
  \psi_0(y) \nonumber \\
\!\!\!\!\!\!&&\!\! + \frac{Q^2}{m_{\rho}^2}  \int^{1}_{\bar{x}_0}\exp\left(
            \Ds \frac{m_{\rho}^2}{M^2}- \frac{Q^2}{M^2}\frac{\bar{x}}{x}
            \right)\frac{dx}{x}
            \Delta_1(0,\bar{x})
            \mathcal{T}^{(1)}(\bar{x},y) \underset{y}\otimes\psi_0(y)+ O(\mathbb{A}_2)
            \Bigg\}, \label{eq:38b}
\end{eqnarray}
\begin{eqnarray}
\!\!\!\!\!\!Q^2 F^{\gamma\pi}_{\text{LCSR};n}\left(Q^2\right)
\!\!\!\!\!&=&\!\!\!\!\!
\frac{N_\text{T}}{a_s^{\nu_n}(\mu^2)}\Bigg\{
 \left(\frac{\mathbb{A}_{\nu_n}(0,s_0;x)}{x}\right)
\underset{x}{\otimes}\psi_n(x)
  +\left(\frac{\mathbb{A}_{1+{\nu_n}}(0,s_0;x)}{x}\right) \nonumber \\
&& \underset{x}{\otimes}
  \mathcal{T}^{(1)}(x,y)
\underset{y}{\otimes}
  \psi_n(y)\,  \label{eq:38c} \nonumber\\
\!\!\!\!\!\!&&\!\! + \frac{Q^2}{m_{\rho}^2}\int^{1}_{\bar{x}_0} \exp\left(
                                   \Ds \frac{m_{\rho}^2}{M^2}- \frac{Q^2}{M^2}\frac{\bar{x}}{x}
                             \right)
                                   \frac{dx}{x} \nonumber \\
                             &&  \times   \bigg[ \Delta_{\nu_n}(0,\bar{x})\psi_n(x)+ \Delta_{1+\nu_n}(0,\bar{x})
             \mathcal{T}^{(1)}(\bar{x},y) \underset{y}\otimes\psi_n(y)\bigg]
             + O(\mathbb{A}_2) \Bigg\}\, .
\label{eq:38d}
\end{eqnarray}
\end{subequations}
%Eq (20a), (20b), (20c)
In Eq.\ (\ref{eq:38a}) we have included in the zero-harmonic spectral density
$\bar{\rho}_0$ the contributions stemming from the twist-four and twist-six terms.
The latter term was first derived in \cite{Agaev:2010aq} and reads
\begin{eqnarray}
  \bar{\rho}_0(Q^2,x)
&=&\!\!
  \psi_0(x)+\frac{\delta_\text{tw-4}^2(Q^2)}{Q^2}x\frac{d}{dx}\varphi^{(4)}(x)+ \bar{\rho}_{\text{tw-6}}(Q^2,x) \, , \\
  ~ \varphi^{(4)}(x)
&=&\!\!\frac{80}3 x^2(1-x)^2\,; \delta_\text{tw-4}^2(Q^2)
   = \left[\frac{a_s(Q^2)}{a_s(\mu^2_0)}
     \right]^\frac{\gamma_\text{tw-4}}{\beta_0}\!\!\delta_\text{tw-4}^2(\mu^2_0),\, \gamma_\text{tw-4} = 32/9\, ,\nonumber \\
   \bar{\rho}_\text{tw-6}(Q^{2}\!,x)  &=&
    8\pi
    \frac{  \alpha_s\langle\bar{q} q\rangle^2 }{f_\pi^2}\frac{C_F}{N_c}\frac{x}{Q^4}
    \left[
        \!-\!
        \left(\frac{1}{1-x}\right)_+
        \!+\!\left(2\delta(\bar{x})-4 x\right)\!+\!
        x\left(
         3+2\ln(x\bar{x})
               \right)
    \right] \, .\label{eq:tw-6} 
\end{eqnarray}
%Eq (21)
Let us conclude this discussion by making two important remarks
with regard to the TFF from the FAPT/LCSRs in (\ref{eq:finalFAPTLCSR}):
(i) Any N$^2$LO contribution to Eqs.\ (\ref{eq:38b}), (\ref{eq:38d}) of order $O(\mathbb{A}_{2+\nu})$
is expected to be sufficiently small due to the reason that $\A_2, \M_2$ are
in the domain $Q^2 \lesssim 1$ GeV$^2$ one order of magnitude smaller than $\A_1, \M_1$ \cite{Bakulev:2006ex}.
(ii) For the numerical calculations of the pion-photon TFF to follow, we replace the
$\delta$-model of the resonances in (\ref{eq:finalFAPTLCSR}) with a more realistic Breit-Wigner model
\cite{Khodjamirian:1997tk,Mikhailov:2009kf}.

\subsection{Numerical results for the TFF and comparison with the experimental data}
\begin{figure}[h]
\hspace{0mm}\includegraphics[width=0.9\textwidth]{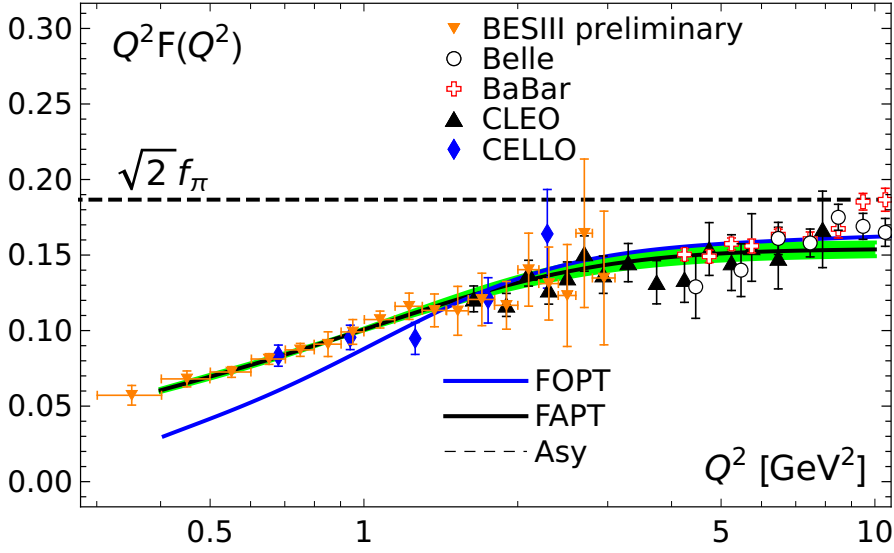}
\caption{
\label{fig:pionFF-strips}
The solid black line and the green strip around it are FAPT predictions for
$Q^2F^{\gamma\pi}_\text{FAPT/LCSR}$, whereas the blue line denotes the
FOPT prediction for $Q^2F^{\gamma\pi}_\text{FOPT/LCSR}$ at the N$^2$LO.
The experimental data of different collaborations are shown in the upper part of the figure.
The single fitted parameter is the scale of the twist-six parameter fixed to the value
$\alpha_s\langle \bar{q}q \rangle^2$ defined at its upper bound
$\langle \bar{q}q \rangle^2=(0.25)^6$ GeV$^6$.
}
\end{figure}

We show in Fig.\ \ref{fig:pionFF-strips} predictions for the TFF
based on (\ref{eq:37}), obtained in two different approaches to include
the QCD radiative corrections using the LCSR method: \\
1) FAPT/LCSRs from Eq.\ (\ref{eq:finalFAPTLCSR})---black curve. \\
2) FOPT/LCSRs from the results in \cite{Mikhailov:2016klg}---blue curve.\\
In both cases we employ the family of the bimodal BMS pion DAs obtained
in \cite{Bakulev:2001pa}.
For their parametrization it is sufficient to employ in the Gegenbauer
decomposition (\ref{eq:gegen-exp}) the coefficients $\{1, a_2,  a_4\}$
derived and discussed in
\cite{Bakulev:2001pa,Stefanis:2014yha,Mikhailov:2016klg}.
The shown predictions are calculated using the coefficient values
at the normalization scale
$\mu^2 \approx 1~\text{GeV}^2$
\cite{Bakulev:2002uc, Bakulev:2004mc}, viz.,
$
 \{a_2(\mu^2)
=
 0.20(+0.05/\!-0.06),a_4(\mu^2)
=
 -0.14(+0.09/\!-0.07),\ldots \}$\footnote{
The values $a_2$ and $a_4$ are strongly correlated approximately along the line
$a_2+a_4=$~const.}.
The other LCSR parameters have been fixed in previous investigations
\cite{Khodjamirian:1997tk,Mikhailov:2016klg} to be
$s_0 \approx 1.5~\text{GeV}^2$,
$M^2 \approx 0.9$~GeV$^2$, $m_{\rho}^2 \approx 0.6$~GeV$^2$,
$\Lambda_{(4)}^{(1-{\rm loop})} \approx 0.3$~GeV ,
$\delta_\text{tw-4}^2(\mu^2)
\approx
 \lambda^2_q/2
\approx
 0.19~\text{GeV}^2$ and are not varied here.
On the other hand, the scale of the twist-six contribution in Eq.\ (\ref{eq:tw-6}),
for both used schemes FAPT/LCSRs and FOPT/LCSRs is fixed at the admissible upper limit
of the condensate $\langle \bar{q}q \rangle^2=(0.25)^6$ GeV$^6$, see, e.g., \cite{Agaev:2010aq}.

Using Eq.\ (\ref{eq:37}) and the partial TFF terms
$F^{\gamma\pi}_{\text{LCSR};n}$ from Eqs.\ (\ref{eq:finalFAPTLCSR}), we obtain for
$Q^2F^{\gamma\pi}_\text{FAPT}(Q^2)$ the prediction shown by the solid black line
for the BMS DA in Fig.\ \ref{fig:pionFF-strips}.
The (green) strip enveloping this curve indicates the estimated range of
theoretical variations of the BMS DA in terms of $a_2$ and $a_4$,
while other uncertainties are not considered here.
The blue line in this figure corresponds to the FOPT predictions
$Q^2F^{\gamma\pi}_\text{FOPT}(Q^2)$ taken at the order N$^2$LO$\beta_0$
within the LCSR scheme in \cite{Mikhailov:2016klg}.
Note that the radiative corrections are negative and become too large
in magnitude below about 1 GeV$^2$.
Obviously, the RG summation effect on the radiative corrections to the TFF
provides a good agreement between the FAPT/LCSR TFF predictions and the preliminary
BESIII data \cite{Redmer:2018uew} even in this very low $Q^2$ domain, where
FOPT turns out to be unreliable.
This is remarkable, given that the experimental margin of error is very small
in this range.
The high-$Q^2$ behavior of the TFF within the FAPT/LCSR approach will be considered elsewhere,
while such predictions at the level N$^2$LO within FOPT/LCSRs can be found in
\cite{Stefanis:2019cfn} with emphasis on the QCD asymptotic limit, denoted
in Fig.\ \ref{fig:pionFF-strips} by the dashed horizontal line.
%%%%%%%%%%%%%%%%%%%%%%%%%%%%%%%%%%%%%%%%%%%%%%%%%%%%%%%%%%%%%%%%%%%%%%%

\section{Conclusion}
We considered the lightcone sum-rule description of the pion-photon
transition form factor in combination with the renormalization group
of QCD and compared the obtained TFF predictions with the corresponding
fixed-order results.
We showed that the LCSR method, augmented with the RG summation of radiative
corrections, naturally leads to a version of fractional analytic perturbation
theory that is free of Landau singularities and provides the possibility to
include QCD radiative corrections in a resummed way \cite{Ayala:2018ifo}.
This FAPT/LCSR approach extends the domain of applicability of the QCD calculation
well below 1 GeV$^2$ and amounts to a significantly smaller total contribution
of radiative corrections in this regime relative to a fixed-order calculation.
To ensure the compliance with the correct QCD asymptotics of the form factor,
new boundary conditions on the FAPT couplings at the origin,
$ \A_\nu(Q^2=0)=\M_\nu(Q^2=0)=0,$ for $0 < \nu \leqslant 1$, have to be imposed.
The FAPT/LCSR approach is best-suited for a detailed comparison with the expected
final BESIII data that bear very small errors in the domain below 1 GeV$^2$.
In fact, as one sees from Fig.\ \ref{fig:pionFF-strips}, the TFF calculated with the
family of endpoint-suppressed BMS DAs \cite{Bakulev:2001pa}, already agrees
with the preliminary BESIII data very well.
On the other hand, the TFF results for the BMS DAs within the FOPT/LCSR scheme
agree with all data compatible with scaling at high-$Q^2$ values, where radiative
corrections can be reliably computed using FOPT \cite{Stefanis:2019cfn}.
\medskip

\textbf{Acknowledgments}\\
S. V. M. acknowledges support from BelRFFR-JINR, Grant No. F18D-002.
A. V. P. was supported by the Chinese Academy of Sciences,
President's International Fellowship Initiative (PIFI Grant No. 2019PM0036).

\newcommand{\noopsort}[1]{} \newcommand{\printfirst}[2]{#1}
  \newcommand{\singleletter}[1]{#1} \newcommand{\switchargs}[2]{#2#1}

\end{document}